\documentclass[12pt,a4paper]{iopart}
\usepackage{amssymb,graphicx,cite}
\eqnobysec
\begin{document}
 
%LATEX FILE OF MANUSCRIPT
%%%%%%%%%%%%%%%%%%%%%%%%%%%%%%%%%%%%%%%%%%%%%%%%%%%%%%%%%%%%%%%%%%%%

%LATEX file of the manuscript
 
%\documentclass[aps,showpacs,twocolumn,floatfix]{revtex4} 
%\documentclass[eqsecnum,aps,twocolumn,epsf]{revtex4} % PH. REV.
%\usepackage{graphicx}
%\documentclass[eqsecnum,aps,twocolumn]{revtex4} % PH. REV.

%\documentstyle[aps,epsf]{revtex}
%\documentstyle[eqsecnum,aps,epsf]{revtex}
%%% <<< epsf commands in the next two lines >>>
%\newcommand{\postscript}[2] {\setlength{\epsfxsize}{#2\hsize}
%\centerline{\epsfbox{#1}}}

%\documentstyle[eqsecnum,aps,epsf]{revtex} % PH. REV. FINAL FORMAT STYLE
%\documentstyle[aps,epsf]{revtex} % PH. REV. FINAL FORMAT STYLE
%%% <<< epsf commands in the next two lines >>>
%\newcommand{\postscript}[2] {\setlength{\epsfxsize}{#2\hsize}
%\centerline{\epsfbox{#1}}}
 
%\documentstyle[preprint,aps]{revtex}
%\documentstyle[eqsecnum,aps]{revtex}
%\documentstyle[aps]{revtex}
%\renewcommand{\baselinestretch}{.88}
 
%\begin{document}
   
%\twocolumn[\hsize\textwidth\columnwidth\hsize\csname@twocolumnfalse\endcsname
 
\title[Bound states of  attractive Bose-Einstein condensates in shallow
traps]{Bound states of attractive Bose-Einstein condensates in shallow
traps in
two and three dimensions }
\author{Sadhan K. Adhikari}
%\affiliation{Instituto de F\'{\i}sica Te\'orica, Universidade Estadual
%Paulista, 01.405-900 S\~ao Paulo, S\~ao Paulo, Brazil}
\address{Instituto de F\'{\i}sica Te\'orica, Universidade Estadual
Paulista, \\  01.405-900 S\~ao Paulo, S\~ao Paulo, Brazil}

\date{\today}
%\maketitle

\begin{abstract}

Using variational and numerical solutions of the mean-field
Gross-Pitaevskii equation for attractive interaction (with cubic or Kerr
nonlinearity) we show that a stable bound state can appear in a
Bose-Einstein condensate (BEC)  in a localized exponentially-screened
radially-symmetric harmonic potential well in two and three dimensions. We
also consider an axially-symmetric configuration with zero axial trap and
a exponentially-screened radial trap so that the resulting bound state can
freely move along the axial direction like a soliton. The binding of the
present states in shallow wells is mostly due to the nonlinear interaction
with the trap playing a minor role. Hence these BEC states are more
suitable to study the effect of the nonlinear force on the dynamics. We
illustrate the highly nonlinear nature of breathing oscillation of these
states. Such bound states could be created in BECs and studied in the
laboratory with present knowhow.

\end{abstract}
\maketitle
 
\section{Introduction}

Solitons are  solutions of  wave equation where
localization is obtained due to a nonlinear
attractive interaction. Solitons
 have been noted in  optics \cite{0},
high-energy physics and water waves \cite{1}, and more recently in
Bose-Einstein
condensates (BEC) \cite{2,3}. The 
bright
solitons of attractive BEC represent  local maxima \cite{3,4a0,4a2,4a1},
whereas 
dark solitons of repulsive BEC
represent local
minima \cite{2}.  

 A
classic  soliton appears  in the  following
nonlinear  Schr\"odinger equation (NLS) in one dimension (1D) in 
dimensionless units \cite{1,7}
\begin{equation} \label{1}
\left[-i
 \frac{\partial }{\partial t}
-  \frac    {\partial^2 }{\partial y^2} -
| \Psi(y,t)|^2 \right]
\Psi(y,t)  =0.
\end{equation} 
The bright solitons of this equation with cubic or Kerr nonlinearity 
are localized solution due to the
attractive
interaction $-| \Psi(y,t)|^2 $
with wave function  at time
$t$ and position
$y$:  $\Psi(y,t)= 
\sqrt{2|{\cal E}| }\exp(-i{\cal E} t){\mbox{sech}}
(y\sqrt
{|\cal E|})$, with $\cal E$ the energy \cite{7}. 
The Schr\"odinger equation
with a nonlinear interaction $-|\Psi|^2$ does not sustain a localized
solitonic solution in two (2D) or three dimensions (3D). However, a
radially-trapped
and
axially-free version
of this equation in 3D does sustain such a bright ``solitonic"
solution \cite{4a0,4a2,4a1} which has been observed experimentally in BEC
\cite{3}.

The solutions  of the NLS (\ref{1}) are bound due to the cubic
nonlinear interaction alone and possesses properties distinct from bound
states in linear potentials \cite{1}. For example, it can travel freely
without distortion and also could be formed at any position in
space. It can  execute breathing oscillation solely under the action
of the nonlinear interaction. The detailed study of this oscillation
should yield information about the nonlinear dynamics. Real stable
bound states of BEC  are not possible in 2D and 3D without
confining traps. However, such states are possible in 2D and 3D employing
a rapidly oscillating nonlinearity \cite{oscn,9} or a rapidly oscillating
dispersion
coefficient \cite{oscd}.  

Nevertheless, it would be  of
great interest to generate BEC bound states in 2D and 3D under stable
conditions 
where the
binding comes mostly from the nonlinear interaction
as 1D. The experimentally observed and theoretically studied
solitonic bound states  of BEC in 3D are created under the action of an
infinite
radial trap in the absence of an axial trap \cite{2,3}. Hence the dynamics
of such a  bound state of BEC 
will
suffer significant distortion due to the infinite radial trap.

We show that it is possible to have stable BEC bound
states
in localized exponentially-screened 
shallow harmonic potential wells in 2D and 3D
bound predominantly due to an attractive   nonlinear
interaction.
As the effect of the  exponentially-screened     potential  on the binding
of the
soliton is expected to be small, the dynamics of these bound states  will
be
mostly controlled by
the  nonlinear
interaction. The formation and the study of such a bound state  could be
of utmost interest in several areas, e.g.,
optics \cite{1}, nonlinear physics \cite{1} and BEC \cite{8}.
We use
both variational as well as numerical solutions of the mean-field
time-dependent quantum-mechanical Gross-Pitaevskii (GP) equation \cite{8}
in our study.

We consider the radially-symmetric configuration in 2D and 3D and the
axially-symmetric configuration in 3D.  In the radially-symmetric case we
employ an  exponentially-screened harmonic  radial trap.  In the
axially-symmetric case the axial
trap is removed and an  exponentially-screened  radial trap is
employed.  Although, the
binding of these objects comes mostly from the nonlinear interaction, in
the radially-symmetric case they are localized in space and cannot
propagate freely. However, in the axially-symmetric configuration as the
axial trap is removed it can move like a soliton in the axial direction.
Unlike a normal trapped BEC, where the breathing oscillation of the
condensate is mostly controlled by the linear trap parameter
\cite{11a,11e},
for the present states, the breathing oscillation is highly nonlinear
in nature. We illustrate this for the radially-symmetric BEC bound states.

Such attractive BEC states could be created in  exponentially-screened  
potential wells in
the
laboratory. 
What is needed is to reduce the height of the confining potential well
in a controlled fashion after a BEC is formed. The experimentalists
routinely perform
such a
reduction in the height of the potential well during the creation of the
BEC by evaporative cooling, by reducing the laser intensity in
optical trapping and/or by reducing the electric current in magnetic 
trapping. Once a BEC is materialized in a shallow trap,
its nonlinear dynamics could be studied in the laboratory and the results
compared with the prediction of the mean-field models.  
This will provide a more stringent test for the mean-field
models.

Apart from interest in the BEC community, the present finding is of
general interest. It is known that, in 3D, an  exponentially-screened
finite harmonic potential  
cannot support a bound state in the linear Schr\"odinger
equation. The state could at best be metastable because of possible
 tunneling through a finite barrier.  
However, the above potential can support a bound state 
in the attractive NLS. A previous
investigation on this topic by Moiseyev {\it et al.}
\cite{mal} led to negative result in 3D and we
comment on it at appropriate places.

In section  2 we present the mean-field model which we use in this  paper. 
In section 3 we develop a time-dependent variational method in 2D and 3D
in the radially-symmetric case. The nonlinear
problem is reduced to an
effective potential well. 
The 
possibility of the appearance of stable
bound states in this potential for a wide range of the
parameters is discussed.  
In section 4 we consider the
complete numerical solution of the GP equation and find the wave
functions of the stable bound states. 
We also illustrate numerically such bound states in the axially-symmetric
configuration in 3D with no axial trap and a weak radial trap. 
Finally,  in section 5 we
present our conclusions.

\section{Mean-Field Model}

\subsection{Radially-Symmetric Well in Two and Three Dimensions}

The usual condensates are formed in a parabolic trap
$V({\cal R})=m^2\omega ^2{\cal R}^2 /2$ where $m$ is the mass of the 
atoms, ${\cal R}$ the
radial distance and $\omega $
the angular frequency of the trap. In the present
investigation, in place of the parabolic trap we consider the
exponentially-screened    
well
$V(r)=m^2\omega ^2{\cal R}^2 \exp(-C {\cal R}^2)/2$, where $C$ is a
positive parameter. 
If $C=0$  one recovers the infinite trap. A small deviation of $C$ 
from
0 leads to a shallow potential well. 

In figure 1 we plot the
scaled 
potential well 
\begin{equation}\label{st}
V(r)= r^2\exp(-c r^2)
\end{equation}
in units of $\hbar \omega$, 
 for $c=0$, 0.05, and 0.1, where $r={\cal R}/l$, $c=Cl^2$ with 
$l=\sqrt{\hbar/(m\omega)}$.  
 Compared
to
the infinite trap for  $c=0$, the two other wells for $c=$ 0.05, and 0.1
are very shallow. There cannot be any stable bound state in these
shallow wells
alone without an attractive nonlinear interaction greater than a critical
value because of the possibility of tunneling to infinity. Such states are
metastable. The height of the potential well has been highly reduced from
infinity to a small finite value $\sim 4$ for $c=0.1$.
The stable attractive bound
state with nonlinearity stronger than a critical value
in such a well 
is bound mostly due to the nonlinear attractive interaction with the
potential well playing a minor role.

\begin{figure}%[!ht]
 
\begin{center}
\includegraphics[width=.75\linewidth]{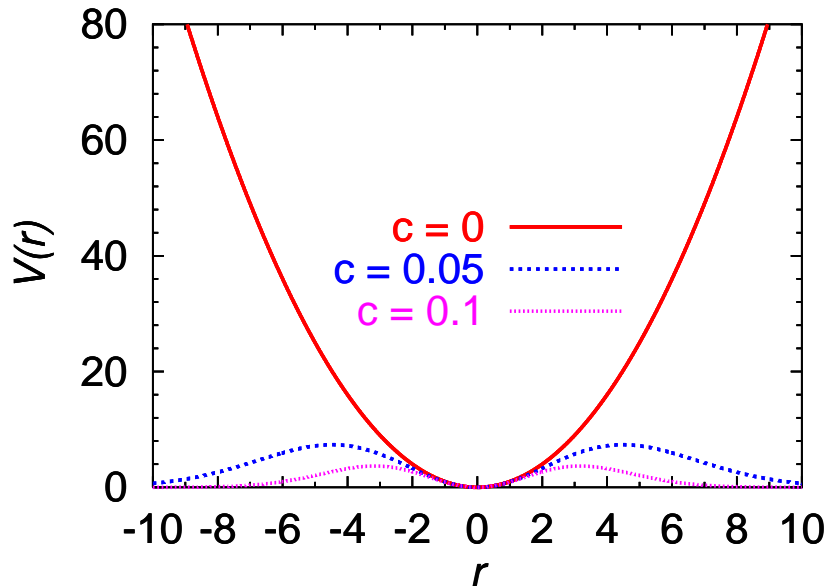}
\end{center}
 
\caption{The scaled potential well  $V(r)=r^2 \exp(-c r^2)$ for $c=0$,
0.05, and 0.1 (upper to lower curves).}
\end{figure}

We use the radially-symmetric time-dependent mean-field
quantum-mechanical GP equation with
attractive
nonlinearity for the present
study \cite{8}. The time-dependent approach is useful to study the
stability of the bound state under small perturbation. 
As we shall not be concerned with a particular
experimental system, we write the GP equation in dimensionless 
variables.   
The GP equation for the 
Bose-Einstein condensate wave
function $\psi({r};t)$ at position ${r}$ and time $t $ can be
written in dimensionless form  as \cite{9}
\begin{eqnarray}\label{d1} 
\biggr[&-& i\frac{\partial
}{\partial t} -{\frac{\partial ^2}{\partial r^2} - \frac{D-1
}{r}\frac{\partial }{\partial r}   }
+
r^2 e^{- c r^2} -
  g\left|
{\psi({ r};t)}\right|^2
 \biggr]\psi({  r};t)=0, 
\end{eqnarray}
where $-g$ is the attractive   nonlinearity and  $D=2, 3,$ is the spatial
dimension. In this paper we shall consider only attractive nonlinearity
corresponding to positive $g$ values. 
Here
length and  time  are expressed in units of
$l$ and  $(\omega/2)^{-1}$, respectively.

In 3D, in terms of number of atoms
${\cal N}$ and atomic scattering length $a$, the nonlinearity is given by 
$g=8\pi  {\cal N}a/l.$ A scaled nonlinearity $n$
is often defined by $n = g/(8\pi)=  {\cal N}a/l.$
The normalization condition in   (\ref{d1}) is
$ \int d^D{ r} |\psi({r};t)|^2 = 1,$ with $d^2r=2\pi r dr $ and 
$d^3r= 4\pi r^2 dr$.

\subsection{Axially-Symmetric Well}

We 
consider the time-dependent GP
equation in dimensionless variables for an axially-symmetric well. 
The GP equation for the Bose-Einstein condensate wave
function $\psi({r},z;t)\equiv \phi({r},z;t)/r$ at radial
position $r$, axial position $z$ and time $t $ can be
written   as \cite{9a}
\begin{eqnarray}\label{d2} 
\biggr[&-& i\frac{\partial
}{\partial t} -{\frac{\partial ^2}{\partial r^2} +\frac{1}{r}
\frac{\partial }{\partial r}- \frac{\partial ^2}{\partial z^2}
  }
+
(r^2+\lambda^2 z^2) e^{- c (r^2+\lambda^2 z^2)}
 \nonumber \\ &-&
\frac{1}{r^2}-  8\pi n\left|\frac
{\phi({ r,z};t)}{r}\right|^2
 \biggr]\phi({  r,z};t)=0, 
\end{eqnarray}
with normalization 
\begin{equation}
2\pi \int_{-\infty}^{\infty}dz \int_0^\infty r dr 
|\psi(r,z;t)|^2=1, 
\end{equation}
where $V(r,z)=(r^2+\lambda^2 z^2) e^{- c
(r^2+\lambda^2 z^2)}
$ is the axial trap.
Here 
length and time are expressed in units of $l (\equiv \sqrt{
\hbar/(m\omega)})$ and $(\omega
/2)^{-1}$. respectively, with $\omega $ the radial trap frequency
and  
$n={\cal N}a/l$  the scaled nonlinearity. 
In our calculation we shall set $\lambda=0$ so that the bound state
develops solitonic transportation property along $z$
axis, and also take a finite  $c$ so that the infinite radial
trap is reduced to a shallow trap.  Hence this model allows the study of 
bound states in 3D under the action of a weak radial trap alone where the
binding comes mostly from the nonlinear interaction.

\section{Variational  Results}

To understand how the bound states are formed in a radially-symmetric
shallow
well in 2D and 3D
 we employ a variational method with the following   
Gaussian
wave function for the solution of  (\ref{d1})  
\cite{9a}
\begin{equation}\label{twf}
\psi(r,t)=  A(t)\exp\left[-\frac{r^2}{2R^2(t)}
+\frac{i}{2}{ \beta(t) }r^2+i\alpha(t) 
\right],
\end{equation}
where $A(t)$,  $R(t)$, $\beta(t)$, and $\alpha(t)$ are the
normalization, width, chirp, and
phase of the soliton, respectively.  In 3D
$A(t)=[\pi^{3/4}R^{3/2}(t)]^{-1}$, and in 2D
$A(t)=[\pi^{1/2}R(t)]^{-1}$.  The Lagrangian density for
generating  (\ref{d1})   is \cite{9a,abdul}
\begin{eqnarray}
{\cal L}(\psi)&=&\frac{i}{2}\left(\dot \psi \psi^*
- \dot \psi^* \psi 
\right)-
\left|\frac{\partial
 \psi}{\partial r} \right|^2  -
r^2e^{-cr^2}|\psi|^2
+\frac{1}{2} g | \psi|^4 ,
\end{eqnarray} 
where  the
overhead dot represents time derivative.
The trial wave function (\ref{twf}) is
substituted in the Lagrangian density and the 
effective Lagrangian $L_{\mbox{eff}}$
is calculated by
integrating the Lagrangian density: $L_{\mbox{eff}}= \int {\cal
L}(\psi)
d ^D r.$ The Euler-Lagrange equations for this effective lagrangian are
given by
 \begin{equation}
\frac{d}{d t}\frac{\partial L_{\mbox{eff}}}{\partial\dot \gamma(t)}=
\frac{\partial L_{\mbox{eff}}}{\partial \gamma(t)},
\end{equation}
where $\gamma(t)$ stands for  $R(t)$, $\beta(t)$, or $\alpha(t)$.

\subsection{Radial symmetry in Three Dimensions}

In 3D the effective Lagrangian is given by \cite{abdul}
\begin{eqnarray}
&L_{\mbox{eff}}&=\frac{\pi^{3/2}A^2(t)R^3(t)}{2}\biggr[-\frac{3}{2}\dot
\beta(t)
R^2(t)+\frac{1}{2\sqrt 2} g A^2(t)\nonumber
\\
&-&2\dot \alpha(t) -\frac{3}{R^2(t)}-3\beta^2(t) R^2(t)-
\frac{3R ^2(t)}{[1+cR^2(t)]^{5/2}} \biggr].
\end{eqnarray}
The Euler-Lagrange equations for $\alpha(t)$, $A(t), \beta(t),$ and
$R(t) $
are  given, respectively,  by
\begin{equation}\label{e1}
\pi^{3/2}A^2R^3= \mbox{constant}=1,
\end{equation}
\begin{eqnarray}\label{e2}    
\dot \beta+\frac{4\dot
\alpha}{3R^2}+\frac{2}{R^4}+2\beta^2+\frac{2}{(1+cR^2)^{5/2}}=
\frac{2gA^2}{3\sqrt
2R^2}, \end{eqnarray}
\begin{eqnarray}\label{e3}    
\dot R= 2 R \beta,
\end{eqnarray}
\begin{eqnarray}\label{e4}    
5\dot \beta+\frac{4\dot
\alpha}{R^2}+\frac{2}{R^4}+10\beta^2+\frac{10}{(1+cR^2)^{7/2}}= 
\frac{gA^2}{\sqrt
2R^2},
\end{eqnarray}
where the time dependence of the different observables is suppressed.
Eliminating $\alpha$ between  (\ref{e2}) and (\ref{e4}) one obtains
\begin{eqnarray}\label{e5}       
2\dot \beta= \frac{4}{R^4}-4\beta^2- \frac{gA^2}{\sqrt
2R^2}-\frac{4-6cR^2}{(1+cR^2)^{7/2}}.
\end{eqnarray} 

From  (\ref{e3}) and  (\ref{e5}) we get the following second-order
differential equation for the evolution of the width
\begin{eqnarray}\label{el}\frac{d^2R}{dt^2}
&=&\frac{4}{R^3}-\frac{g}{\sqrt{2
\pi^3}}\frac{1}{R^4}-\frac{4R-6cR^3}{(1+cR^2)^{7/2}}\\
&=& -\frac{d}{dR}\left[\frac{2}{R^2}- \frac{g}{3\sqrt{2
\pi^3}}\frac{1}{R^3}+ \frac{2R^2}{(1+cR^2)^{5/2}} 
\right]. \label{el2} 
\end{eqnarray}
The quantity in the square brackets of  (\ref{el2}) is the effective
potential $U(R)$ of the  equation of motion:
\begin{equation}\label{ef1}
U(R)=\frac{2}{R^2}- \frac{8}{3\sqrt {2\pi}}\frac{n}{R^3}+
\frac{2R^2}{(1+cR^2)^{5/2}}.
\end{equation}
Small oscillation around a
stable configuration might be  possible when there is a minimum in this
effective
potential.

\begin{figure}%[!ht]
 
\begin{center}
\includegraphics[width=.49\linewidth]{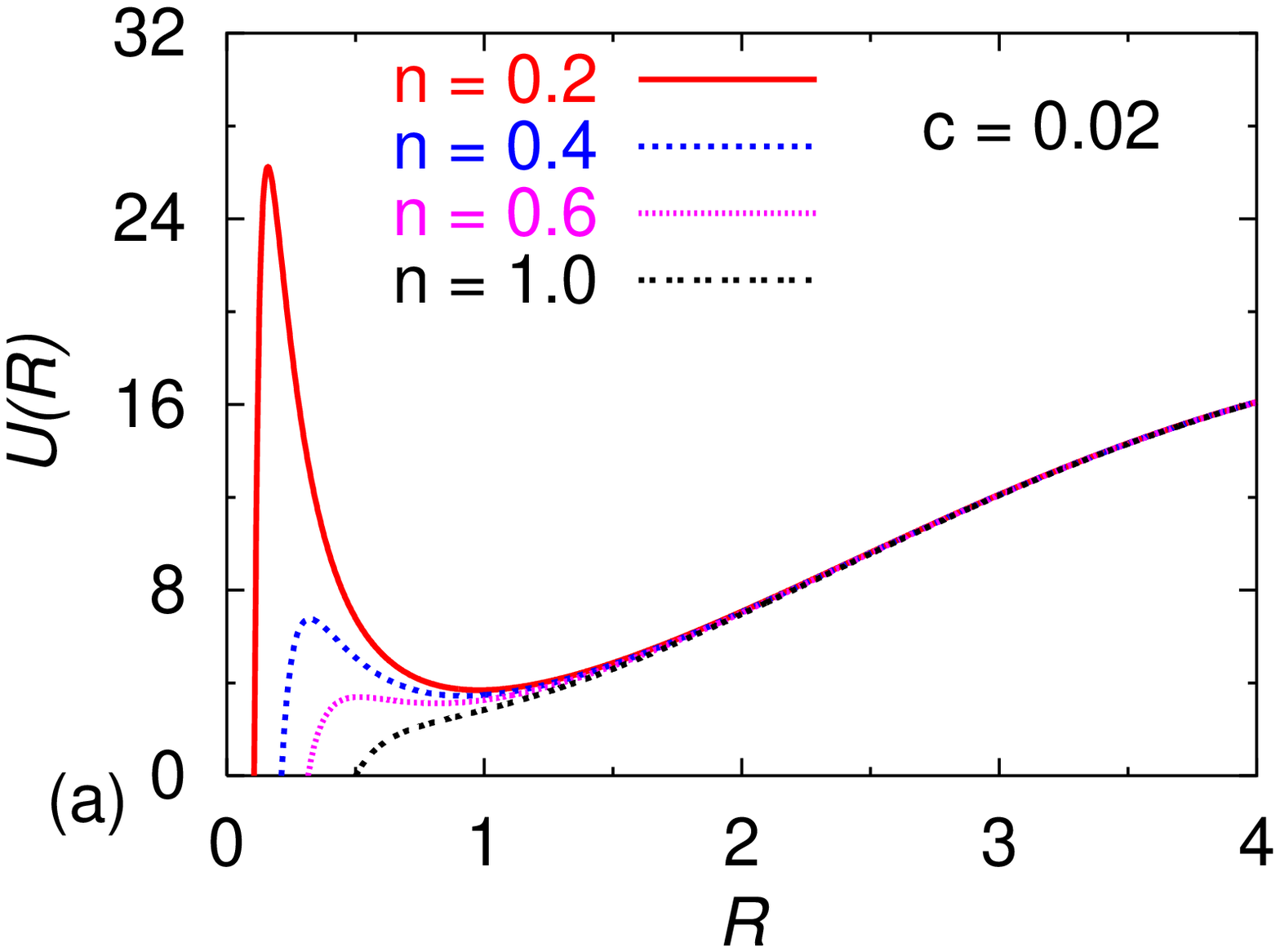}
\includegraphics[width=.49\linewidth]{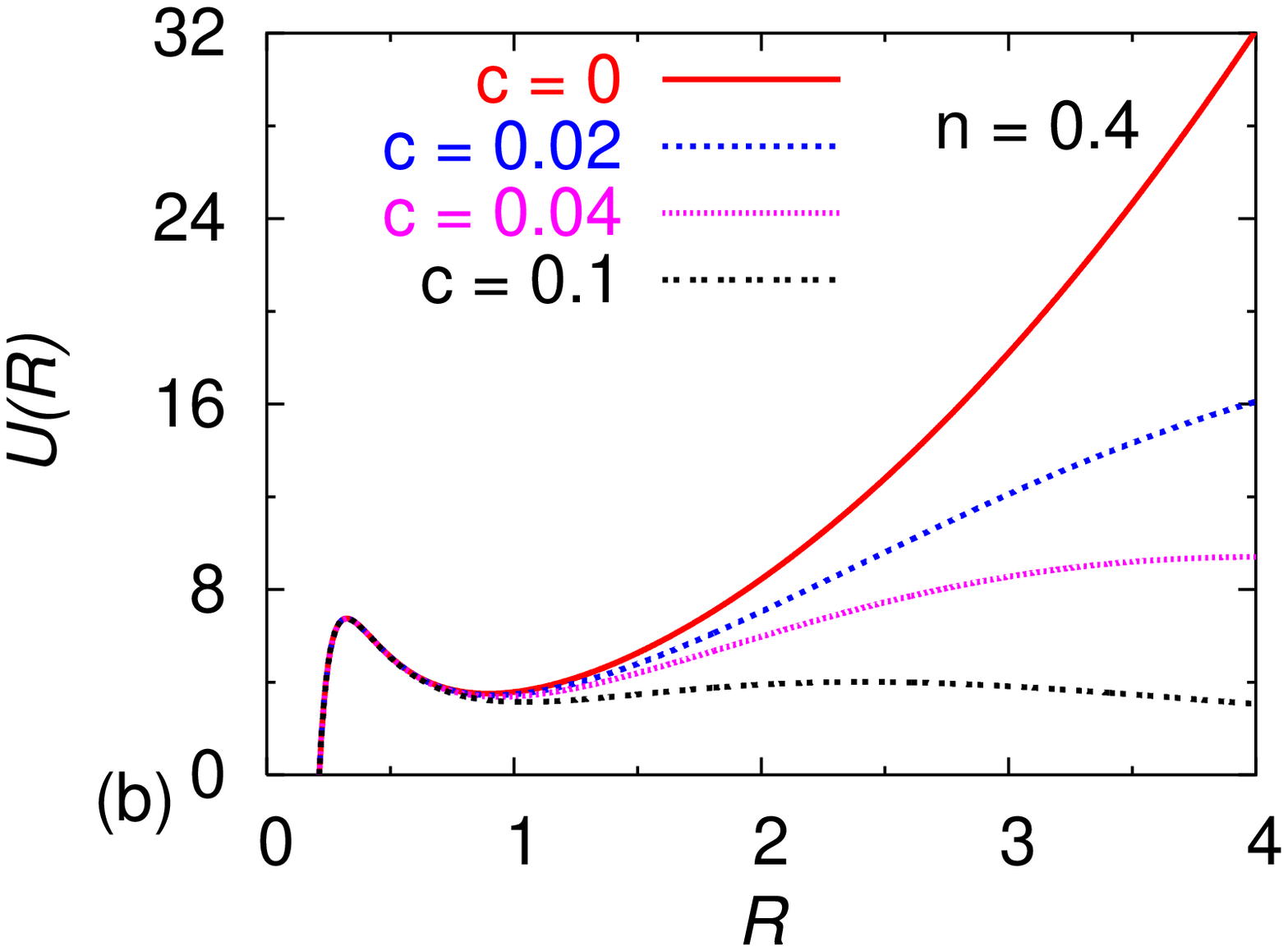}
\end{center}
 
\caption{The effective potential $U(R)$ in 3D of
 (\ref{ef1}) vs. 
$R$  in
arbitrary units for (a) $c=0.02$ and $n\equiv g/(8\pi)=0.2,0.4, 0.6,$ and
$1.0$ 
(upper to
lower curves) and for 
(a) $n=0.4$ and $c=0,0.02, 0.04,$ and $0.1$
(upper to
lower curves).
}
\end{figure}

A study of the effective potential (\ref{ef1}) as a function of parameters
$n$ and $c$ reveals some interesting features. The small-$R$ part of
$U(R)$ is insensitive to $c$ whereas, for small $c$, the long-$R$ part of
$U(R)$ is insensitive to $n$. This is illustrated through the plot of
$U(R)$ vs. $R$ for different $n$ and $c$ in figures 2 (a) and (b). The
minimum of $U(R)$ for $R\approx 1$ representing a shallow effective well
can allow a bound solution of the GP equation. For $c>0$, there cannot be
a stable bound state in the linear problem where $g=n=0$.  Although a
minimum could appear in $U(R)$ in that case for a small $c$, the state
becomes metastable and decays 
eventually by tunneling to infinity. This is possible as $U(R) \to 0$ as
$R\to \infty$.
The variational approach
alone does not distinguish between a stable and a metastable state. 

In figure 2 (a) we see that as
the attractive nonlinearity $n$ is increased for a fixed $c$, one of the
walls of the well is gradually lowered and for a sufficiently large $n$
this wall is completely absent and the condensate collapses into the
infinitely deep well at the center. This happens for critical nonlinearity
$n_{\mbox{crit}}\approx 0.6$ representing the onset of collapse. This
value is to be compared with the accurate numerical value of 0.575
\cite{8,9a} for the infinite harmonic potential with $c=0$. We also find
in figure 2 (b)  that, for a fixed $n$, as $c$ is increased from 0, the
shallow well lasts up to a finite value of $c$ beyond which there is no
well for the formation of a bound state $-$ stable or metastable.  The
simple variational study
qualitatively describes the formation of an attractive BEC in a shallow
finite potential. Also, in such case of critical equilibrium the
variational treatment usually over-binds the system and does not yield
reliable results, for which we consider the direct numerical treatment in
the next section.

\subsection{Radial symmetry in Two Dimensions}
In 2D the effective Lagrangian is given by \cite{abdul}
\begin{eqnarray}
&L_{\mbox{eff}}&={\pi A^2(t)R^2(t)}\biggr[-\frac{1}{2}\dot
\beta(t)
R^2(t)+\frac{1}{4} g A^2(t) -\dot \alpha(t)\nonumber
\\
&-& \frac{1}{R^2(t)}-\beta^2(t) R^2(t)-
\frac{R ^2(t)}{[1+cR^2(t)]^{2}} \biggr].
\end{eqnarray}
As in the three-dimensional case one can write the 
the coupled Euler-Lagrange equations for $\alpha(t)$, $A(t), \beta(t),$
and
$R(t) $. After some algebra one can eliminate the variables  $\alpha(t)$,
$A(t),$ and $ \beta(t)$  from these equations and 
\begin{figure}%[!ht]
 
\begin{center}
\includegraphics[width=.7\linewidth]{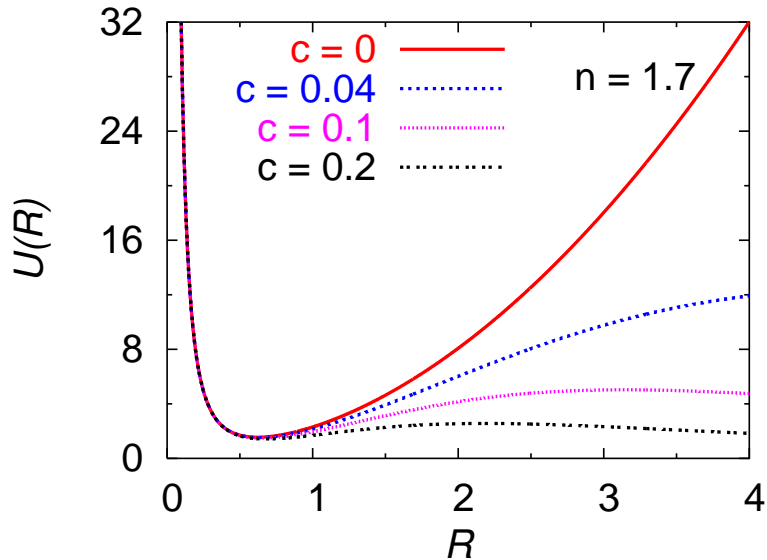}
\end{center}

\caption{The effective potential $U(R)$ in 2D of
 (\ref{ef2}) vs.
$R$  in
arbitrary units for  $n\equiv g/(2\pi)=1.7$ and $c=0,0.04, 0.1,$ and $0.2$
(upper to
lower curves).
}

\end{figure}
obtain the following second-order
differential equation for the evolution of the width $R$
\begin{eqnarray}\label{fl}\frac{d^2R}{dt^2}
&=&\frac{4}{R^3}-\frac{g}{\pi
}\frac{1}{R^3}-\frac{4(R-cR^3)}{(1+cR^2)^{3}}\\
&=& -\frac{d}{dR}\left[\frac{2(1-g/4\pi)}{R^2}
+ \frac{2R^2}{(1+cR^2)^{2}}
\right]. \label{el3} 
\end{eqnarray}
In this case
the effective potential $U(R)$ for the evolution of the width 
is given by
\begin{equation}\label{ef2}
U(R)=\frac{2-n}{R^2}+ \frac{2R^2}{(1+cR^2)^{2}},
\end{equation}
where $n=g/(2\pi)$ in 2D.

The  study in 2D reveals similar qualitative features  as in 3D. From
 (\ref{ef2}) we see trivially that the
onset of collapse is given in this case by $n=2$. This has to be compared
with the numerical value $n \approx 1.85$ \cite{col2,col3,col4}
obtained by solving the GP equation in 2D  
with $c=0$.
For $n>2$, $U(R)$ does
not have a barrier for small $R$ and the collapse cannot be avoided. There
is no such simple analytic condition of collapse in 3D where it can be
numerically obtained from
 (\ref{ef1}).

In
figure 3 we plot $U(R)$ for different $c$ values for $n=1.7$. As in 3D we
find
that with an increase of $c$ the outer wall of the well is gradually
reduced and for a large enough $c$ this wall is completely removed and no
bound state is possible. Again, although the simple variational
study
reveals these qualitative features, it is not so useful for a quantitative
study in this region of critical stability.  Also, it cannot distinguish
between a stable and a metastable state.

\section{Numerical  Results}

We solve the GP equations (\ref{d1}) and (\ref{d2}) for shallow potential
wells  in 2D and
3D numerically using
the
split-step time-iteration method employing the Crank-Nicholson
discretization
scheme described recently \cite{11a,11b}. In the radially-symmetric case 
the time iteration is started with the
known oscillator solution of these equations with  $n=c=0$.
In the axially-symmetric case in the initial
configuration a small finite axial trap parameter $\lambda$ is employed. 
Then in the course of time iteration the attractive nonlinearity $n$ and
the  parameter $c$ of the potential are 
switched on slowly. In the axially-symmetric case the parameter $\lambda$
is gradually turned off during time iteration.
If stabilization could be obtained 
for the chosen parameters one already obtains the required bound state 
in the shallow trap.

Stable bound
states are indeed found in all cases for various ranges of
parameters:  $c$ and $n$.
Some plausible properties of the bound states are found. 
Stable bound states are
only formed  for $c$ less than a critical value. For larger $c$, from the
wisdom obtained in variational calculation, the effective potential $U(R)$
does not have a minimum and there cannot be any bound state.  For a fixed
$c$, bound states are found for $\bar  n  <n<n_{\mbox{crit}}$. For
$n>n_{\mbox{crit}}$ the system  is very  attractive and
collapses. For $n<\bar n$, the system is very weakly attractive 
and
expands
to infinity. The quantum state is then metastable. 
The numerical values of $ \bar  n$ and
$n_{\mbox{crit}}$ are different in 2D and 3D. 
As $c$ is reduced to zero, both $\bar n $ and $
n_{\mbox{crit}}$ decreases. Finally, at $c=0$,  $n_{\mbox{crit}}$ reduces
to
the critical nonlinearity for collapse in the full harmonic trap
\cite{8,9a}
and
$\bar n$ reduces to zero.  This is consistent with the fact that in the
full
harmonic trap, one can have a stable bound state for all attractive
nonlinearities smaller than a critical value. 
Also, for $c>0$ the critical
nonlinearity for collapse $n_{\mbox{crit}}$ is larger than the critical
nonlinearity for the full harmonic trap.

\begin{figure}%[!ht]
 
\begin{center}
\includegraphics[width=.7\linewidth]{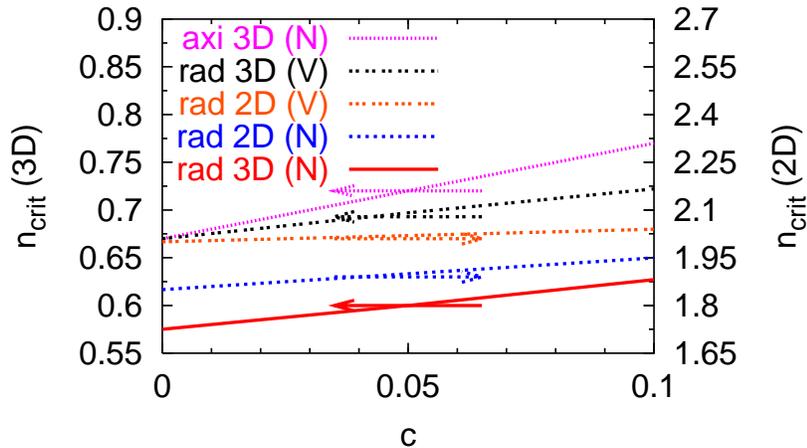}
\end{center}

\caption{Results of variational (V) and numerical (N) 
calculations for 
critical
nonlinearity $n_{\mbox{crit}}$ 
in radially-symmetric 3D, axially-symmetric 3D, 
 and radially-symmetric 2D  for different values of
the parameter $c$. 
}

\end{figure}

\begin{figure}%[!ht]
 
\begin{center}
\includegraphics[width=.7\linewidth]{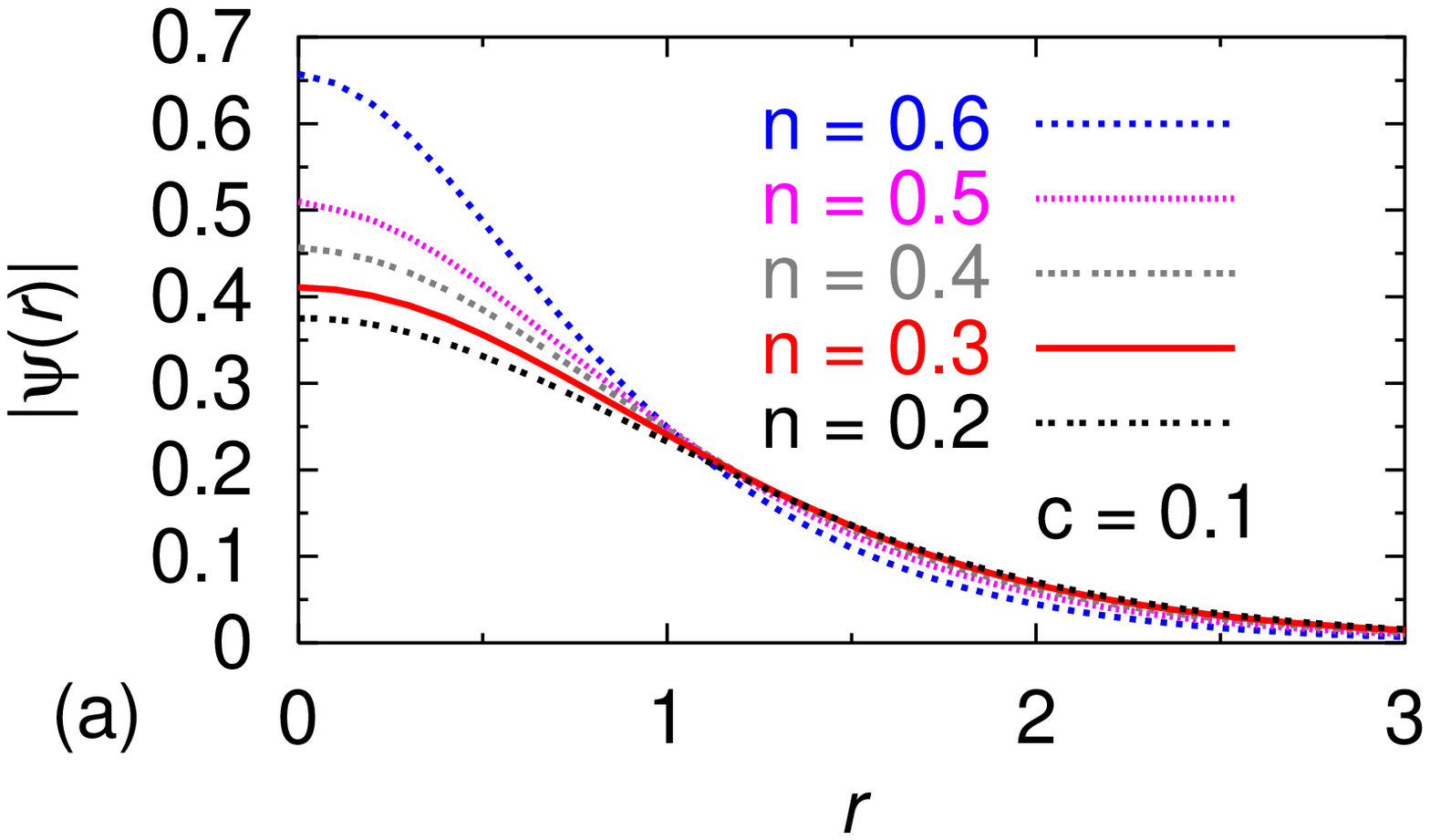}
\includegraphics[width=.7\linewidth]{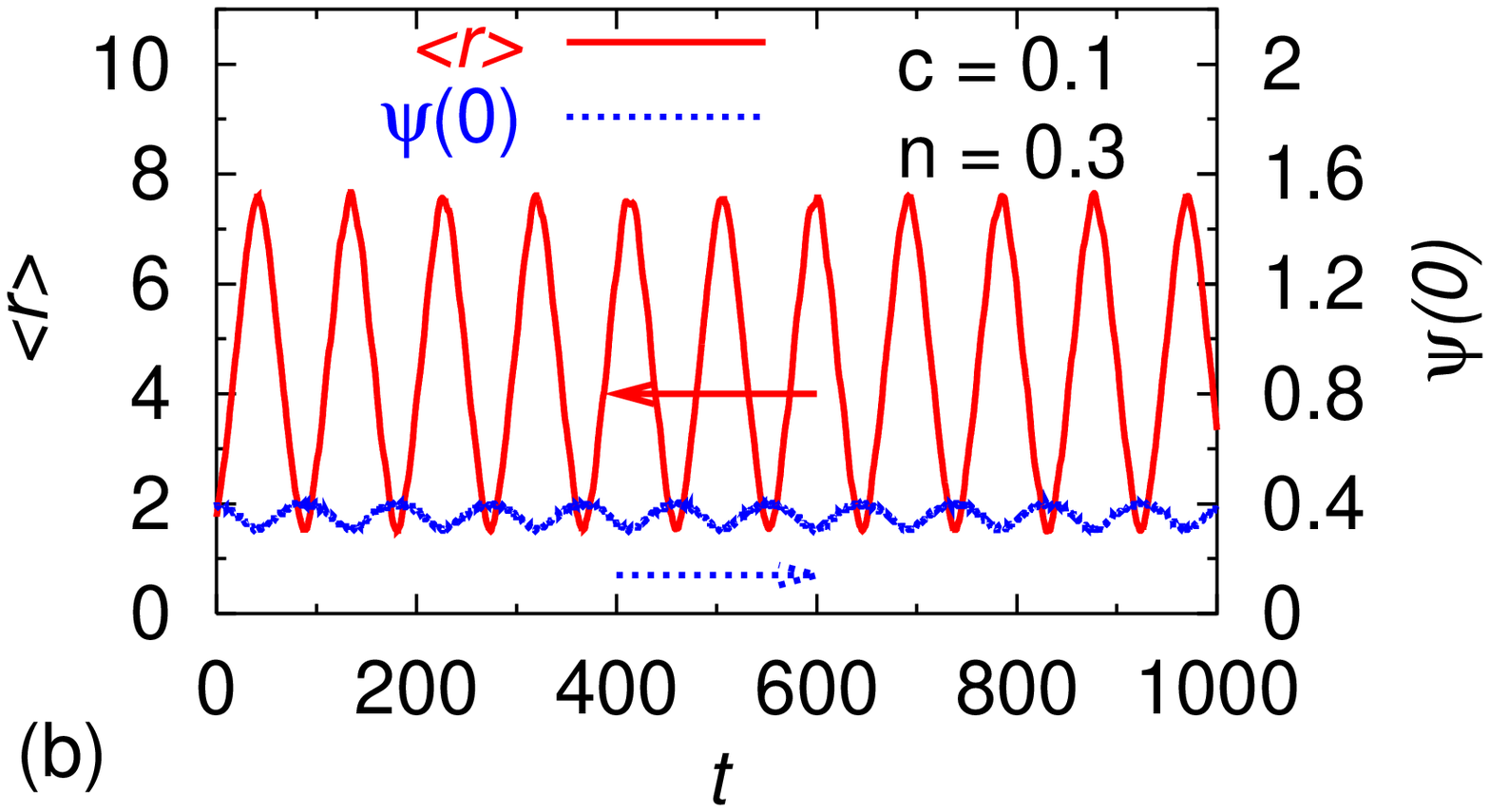}
\end{center}
 
\caption{
(a) Stable bound-state wave function  $|\psi(r)|$ of the 
radially-symmetric GP equation
(\ref{d1})  in 3D
with $c=0.1$
and scaled nonlinearity $n\equiv g/(8\pi) = 0.6, 0.5, 0.4, 0.3,$ and 0.1 
(upper to lower curves);
(b)  $|\psi(r=0)|$ (lower curve) and rms radius $\langle r \rangle$
(upper curve) of the bound-state
wave function  for  $c=0.1$   and $n= 0.3$
executing breathing oscillation initiated by multiplying the shallow
trapping potential suddenly at $t=0$ by 0.95.
}
\end{figure}

\begin{figure}%[!ht]
 
\begin{center}
\includegraphics[width=.75\linewidth]{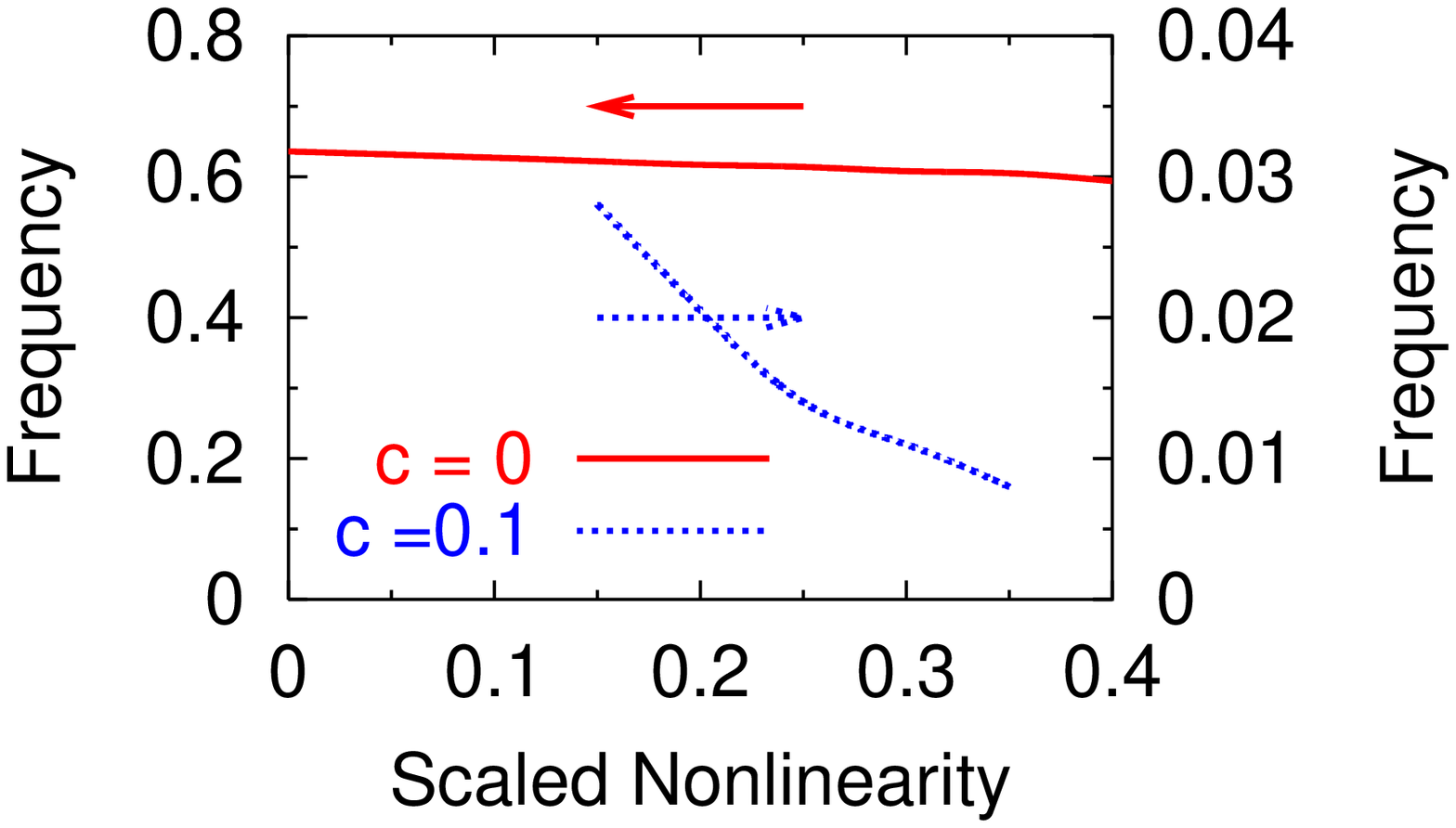}
\end{center}
 
\caption{Frequency of breathing oscillation vs. scaled nonlinearity 
$n$ for $c=0$ and 0.1 in the radially symmetric 3D case. Note the
different scales used for
$c=0$ and 0.1.
}
\end{figure}

\subsection{Radial Symmetry in Three Dimensions}

First we present the numerical results in 3D. 
For $c=0$,  $n= {\cal N}a/l=0.575$ is the critical
nonlinearity for collapse \cite{8}. For  $c=0.1$, the confining shallow
potential
is weaker and 
the critical nonlinearity of
collapse is larger and a stable bound state  could be 
formed for $n =0.6$. As the nonlinearity $n$ is increased, the system
becomes more attractive and the wave function develops a central peaking.
In figure 4 we plot the variational and numerical results for  
the critical nonlinearity for collapse for different 
values of $c$ in 3D and 2D. 
This plot illustrates how  the critical nonlinearity
increases with $c$ for radial and axial (discussed in section
4.3) symmetry in 3D and radial
symmetry in 2D (discussed in section 4.2). The variational 
results are always larger than the numerical ones.

Recently, there has been another study of BEC bound states in shallow
traps (\ref{st})  by Moiseyev {\it et al} \cite{mal}. 
They conclude that there cannot be stable
bound states
in 3D in shallow traps  (\ref{st}). We do not know the reason for this
disagreement. As $c \to 0$, the potential  (\ref{st}) tends to an infinite
harmonic trap. For large $c$ the potential becomes very shallow and it is
reasonable that  
these bound states
should disappear. We find it hard to believe that these states should not
exist for a small value of $c$ when the potential tends to  a pure
harmonic one.  We iterated the GP equation for more than 10 000 time
units corresponding to a million time iterations and the bound-state wave 
function remained  unchanged demonstrating its stability. 

In figure 5 (a) we plot the wave function for the bound state of the GP
equation (\ref{d1})
for $c=0.1$ and different   $n\equiv g/(8\pi)={\cal N}a/l $.
To study the stability of the
bound state  under small perturbation, we study breathing
oscillation in
the bound state by suddenly changing the strength of the shallow trapping
potential from 1 to 0.95. The system executes harmonic oscillations
around a stable configuration illustrated in figure 5 (b), where we plot 
the root-mean-square (rms) radius $\langle r \rangle$  and  $\psi(0)$
vs. time for $c=0.1$ and $n=0.3$. The prolonged stable oscillation of
these
states for an interval of time of more than 1000 units assures us of their
stable, opposed to metastable,  nature.

From figure 1 we see that for $c=0.1$ the shallow trapping potential is
highly reduced compared to the full harmonic potential. In the GP equation
(\ref{d1}), in this case the binding is provided mostly by the nonlinear
interaction.  Hence the
oscillation depicted in figure 5 (b)
is highly nonlinear in nature. 
In the $c=0$ case we 
studied in detail this breathing
oscillation and the frequency $\nu$ of the breathing    oscillation 
was  roughly double the frequency of the trapping
potential $\omega $ (= 2 in present units)
and independent of nonlinearity $n$: 
$\nu \approx 2\times  \omega/2\pi
=2/\pi \approx  0.636$ \cite{11a}. 
However, the present frequency of highly nonlinear oscillation
in figure 5 (b) is approximately 0.011 $-$
completely different from the $c=0$ case. 
The breathing frequency
is found to be  sensitive to the nonlinearity in the GP equation for
$c=0.1$.  To
illustrate this  
we plot in figure 6 the variation in 
breathing frequency with nonlinearity for $c=0$ and 
$c=0.1$. The frequencies for $c=0$ and $c=0.1$ 
are distinct: the frequencies for the shallow well with $c=0.1$ 
are sensitive to the nonlinearity. For this small variation of 
nonlinearity the frequencies for the infinite well are practically 
independent  of nonlinearity. However, the frequencies for the infinite
well will gain a weak dependence on nonlinearity 
for large repulsive interactions (not discussed here). 
As  nonlinearity $n$  is reduced below  0.15 for 
$c=0.1$, first the breathing oscillation enters a 
chaotic   regime with no well defined frequency and 
with further reduction in  $n$ the bound state becomes metastable. Also,
as  nonlinearity $n$  is increased 
beyond 0.35 for $c=0.1$, the oscillation 
becomes irregular and we
did not study it in detail.

The highly nonlinear  breathing oscillation of the condensate discussed
above is not only of theoretical interest but could be studied in the
laboratory with present knowhow. The infinite harmonic trap can slowly
be reduced to generate a shallow potential after the formation of an
attractive condensate and the highly nonlinear breathing oscillation could
be studied. A careful study of this oscillation  will reveal interesting
features of this dynamics and could be used to compare
the experimental results with mean-field predictions.

\subsection{Radial Symmetry in Two Dimensions}

Now we present the numerical results in 2D. 
In figure 7 we plot the wave function for the solution of the GP equation
(\ref{d1}) for $c= 0.1$ and different values of nonlinearity $n=g/(2\pi)$.
For $c=0$ the critical nonlinearity for collapse is $n\approx 1.85$ 
\cite{col2,col3,col4}. For $c=0.1$,
one can have a larger value for the critical nonlinearity. 
A variation of critical nonlinearity in this case is shown in figure 4.
With the increase of the nonlinearity $n$, the nonlinear atomic 
attraction increases and the condensate is confined to a smaller region in
space. Consequently, in figure 7 the wave function has a stronger
central peaking    for larger nonlinearity as in 3D. 
The critical nonlinearity for collapse calculated by 
Moiseyev {\it et 
al} \cite{mal} ($\sim 1.72$)  is typically smaller than our estimate
($> 1.85$). 
[Note that the values reported by them ($\sim 0.86$)  differ from the
present values by
a factor of 2 because of a slightly different nonlinear equation used by
them.] Again the reason for this disagreement is not clear.

We also noted the highly nonlinear breathing oscillation in this case for
$c=0.1$ 
initiated by a sudden change in the strength of the harmonic interaction. 
Unlike in the $c=0$ case, where the frequency of breathing oscillation is
virtually independent of small changes in nonlinearity, in this case 
the  frequency of breathing oscillation is quite  sensitive to the
nonlinearity in the GP equation. However, a complete and comprehensive
analysis of this nonlinear breathing oscillation is beyond the scope of
this paper and will be one of future interest.

\begin{figure}%[!ht]
 
\begin{center}
\includegraphics[width=.75\linewidth]{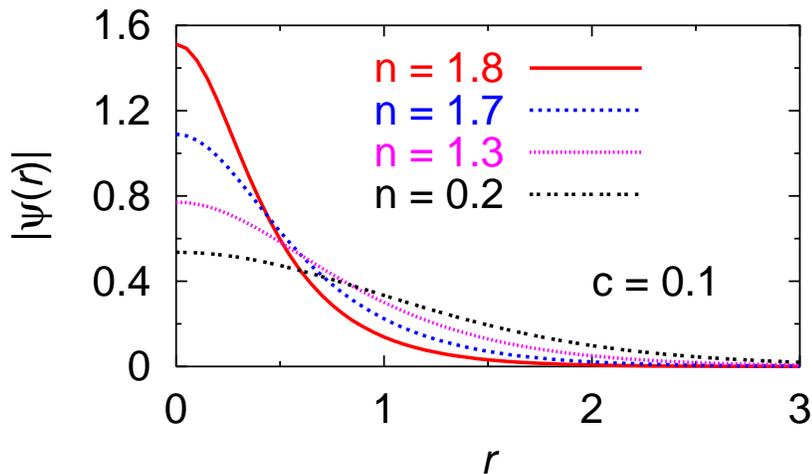}
\end{center}
 
\caption{
 Stable bound-state wave function  $|\psi(r)|$ of the radially-symmetric 
GP equation
(\ref{d1})  in 2D
with $c=0.1$
and nonlinearity $n\equiv g/(2\pi)= 1.8, 1.7, 1.3,$ and  0.2  
(upper to lower curves).
}
\end{figure}

\subsection{Axial Symmetry in Three Dimensions}

Finally, we consider the case of an axially-symmetric stable bound state
in
3D. This
is of interest as most of the experimental condensates possess axial
symmetry. We consider the numerical solution of the time-dependent GP
equation (\ref{d2}) for our purpose, with $\lambda =0$ and $c=0.1$. This
case corresponds to the $z$-independent scaled potential $V(r,z)= r^2 \exp
(-0.1 r^2)$. This means that the bound state is free to move along the
axial
$z$ axis and bound radially by the above weak potential. The critical
nonlinearity for collapse in this case for different $c$ is plotted in
figure 4. The profile of
the bound states in this case is shown in figures 8 (a) $-$ (d) where we
plot
the stationary wave function $|\psi(r,z)|$ of  (\ref{d2}) for
nonlinearities $n=0.1,   0.4,  0.6$ and 0.7, respectively. As
the axial trap is removed the condensate is very long in the $z$ direction
and possesses a small radius. In figures 8 we see that the condensate has
the shape of a cigar about 50 units long with a radius of 2 units. As the
nonlinearity is increased from $n = 0.1$ to 0.7 in figures 8, the
condensate collapses towards the center resulting in a high central
peaking. This aspect is similar to solitons in 1D where a similar peaking
is observed with the increase of nonlinearity.  However, there is no
collapse in 1D.
In the radially
symmetric case $\lambda=1$ and $c=0$, the BEC could be formed for
nonlinearity $n$ less than a critical value 
$n_{\mbox{crit}} =
0.575$ \cite{8,9a}. For the usual three-dimensional bound state 
$c=\lambda=0$,
$n_{\mbox{crit}} = 0.67$ \cite{4a0,4a1}. In the present case $c=0.1$
and $\lambda=0$, $n_{\mbox{crit}}$ is  greater than 0.7. In
all cases, for $n > n_{\mbox{crit}}$, the bound state becomes highly
attractive and collapses, so that no stable bound state  could  be
formed.  The
reduction of $\lambda$ from 1 to 0 or the increase in $c$ from 0 to 0.1 in
these cases denotes a weakening of the confining trap. Consequently, the
condensate can occupy a larger region of space and accommodate a larger
number of particles (or nonlinearity) before collapsing, corresponding to
an increase in $n_{\mbox{crit}} $.
In this case also one could have breathing oscillation as in the
radially-symmetric case. This oscillation will be more complicated,
specially
for the interesting case of $c=0.1$, 
 involving two degrees of freedom and its study is not appropriate for
this paper. 

\begin{figure}%[!ht]
 
\begin{center}
\includegraphics[width=.35\linewidth]{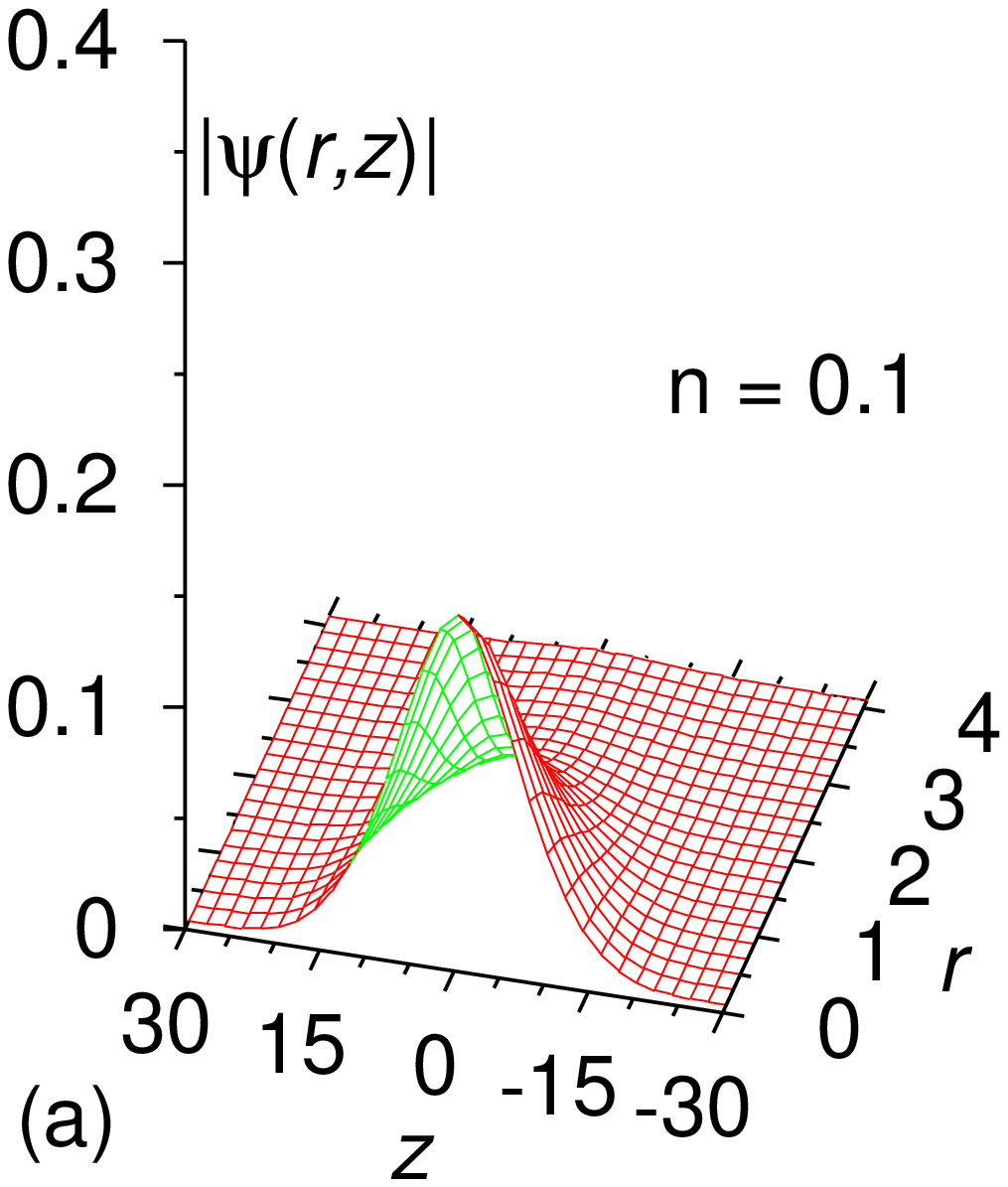}
\includegraphics[width=.35\linewidth]{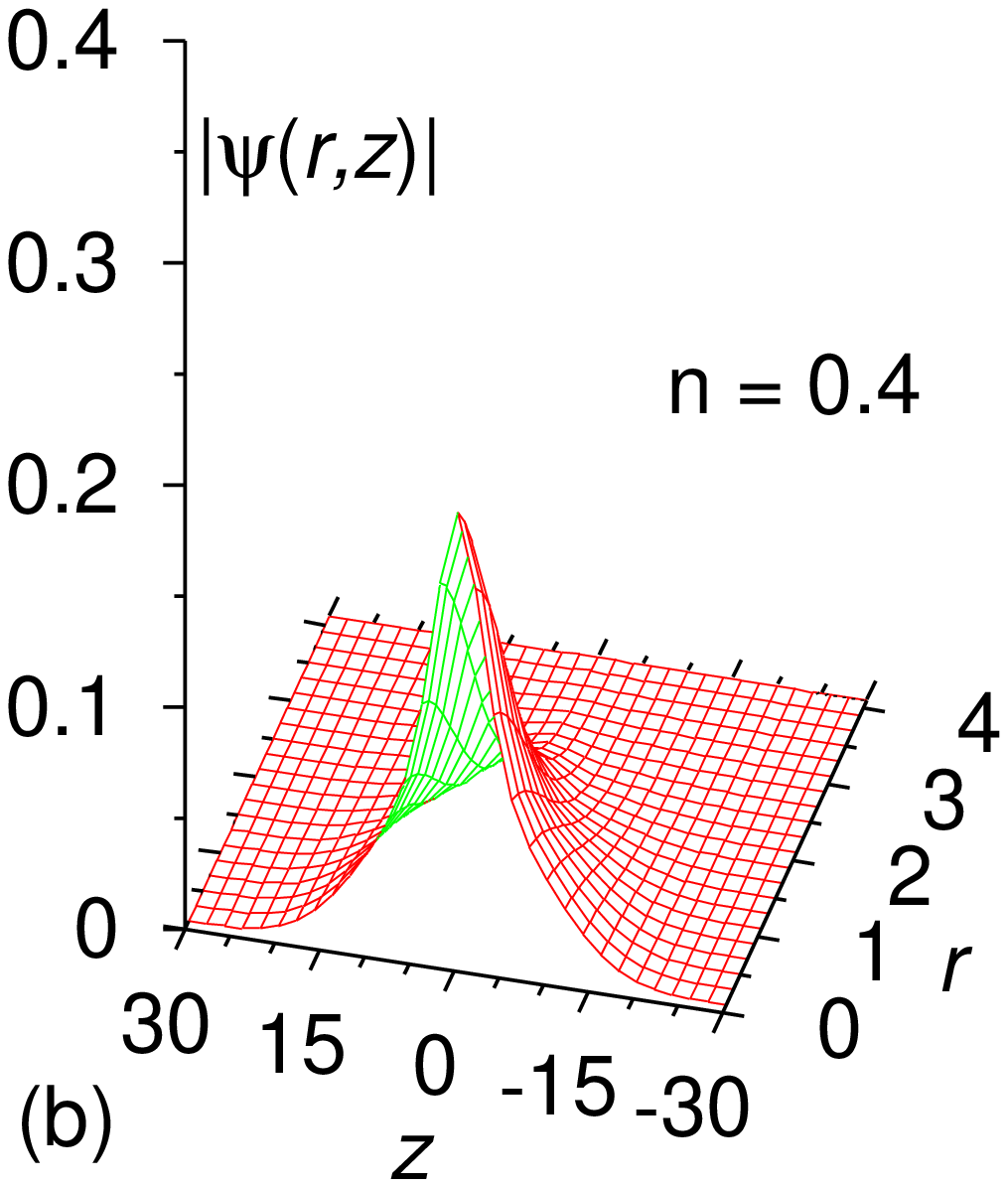}
\includegraphics[width=.35\linewidth]{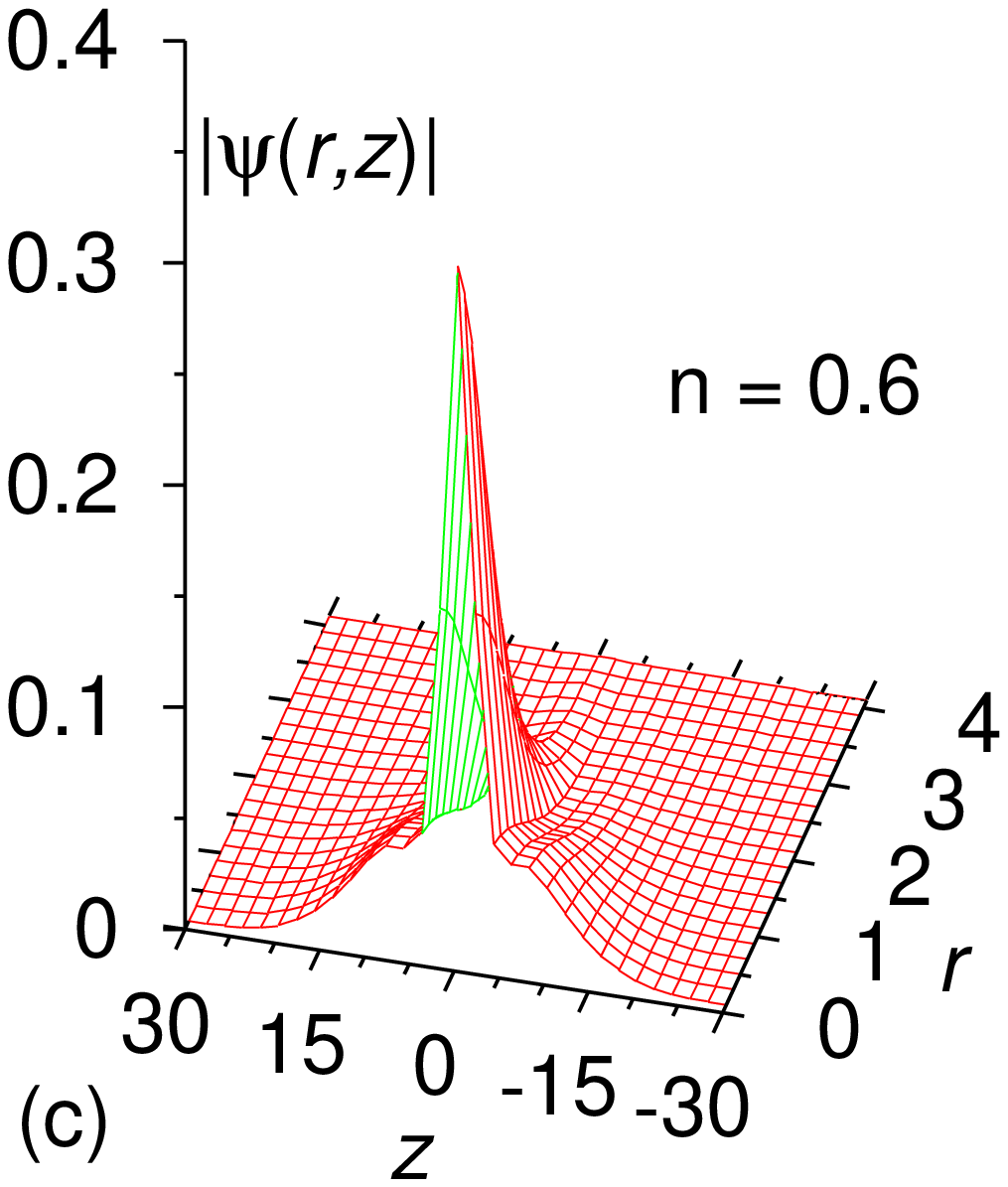}
\includegraphics[width=.35\linewidth]{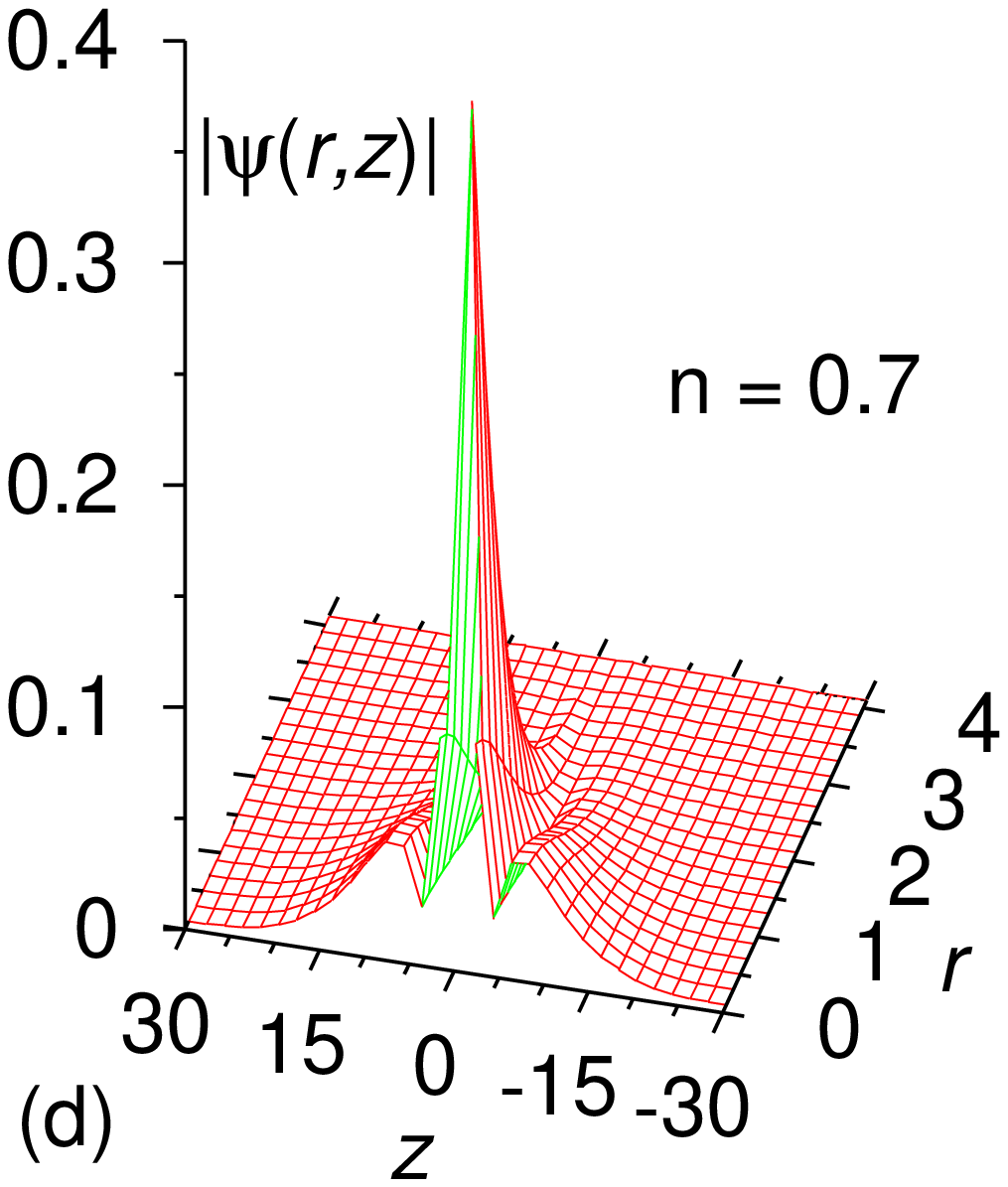}
\end{center}
 
\caption{
 Stable bound-state wave function  $|\psi(r,z)|$ of the 
axially-symmetric GP equation
(\ref{d2})   in 3D
with $c=0.1$ and $\lambda=0$ 
and nonlinearity $n=$  (a)  0.1, (b) 0.4, (c) 0.6 and  (d) 0.7. The
$z$-independent trapping potential is $V(r,z)=r^2
\exp(-0.1r^2)$.}
\end{figure}

\section{Discussion and Conclusion}

  The study of the dynamics of   one-dimensional     solitons is of
great interest in different areas of science and mathematics because
of its intrinsic nonlinear nature. In the
three-dimensional world,   one-dimensional systems can only be achieved in
some approximation. Also, it is difficult to find a
true three-dimensional soliton in nature 
bound only under
the action of a nonlinear interaction. 
BECs with attractive atoms 
are very appropriate for the
study of
nonlinear dynamics as the GP equation describing the BEC dynamics
possesses
a cubic nonlinear (Kerr) interaction  which is identical with the sole 
interaction term in the one-dimensional NLS (\ref{1}). However, unlike in
one-dimensional solitons, in 2D and 3D an infinite  confining  potential
has to
be included in the GP equation for the formation of a stable BEC
\cite{8}. To have
a greater similarity with freely moving one-dimensional solitons, later it
was shown that stable three-dimensional bound states moving in the axial
direction like a soliton  could be formed after removing the axial trap
in an axially-symmetric configuration maintaining  an infinite radial
trap \cite{4a0}. However, the strong infinite radial trap, and not the
weaker
cubic
nonlinear
interaction, controls the essential dynamics of these objects. In this
paper we showed that similar objects could be formed even when the
infinite
radial trap is reduced to an exponentially-screened harmonic 
potential. Under the action of such a
weak radial trap, the cubic nonlinear interaction  plays a major
role on dynamics.

To establish the present suggestion we solved the GP equation numerically
to find the bound states in the radially-symmetric potential 
$V(r)=r^2 \exp(-0.1r^2)$   in 2D and 3D. We studied in some
detail the breathing oscillation of such a bound state  in 3D. The
frequency of
this oscillation is found to be  
sensitive to the nonlinearity of the problem. In the infinite harmonic
potential $V(r)=r^2 $, the frequency of breathing oscillation is found to
be weakly dependent on  the nonlinearity of the problem (at least for 
small nonlinearities) and determined mostly by
the frequency of the trapping potential.  We finally
demonstrate  numerically the
existence of these bound states in the $z$-independent 
axially-symmetric potential $V(r,z)=r^2 \exp(-0.1r^2)$. 
Such axially free and
radially bound states  of BEC under the action of a weak trap
can
be
formed and studied in the laboratory. 

It is appropriate to make 
a prediction of the height of the trap (\ref{st}) to be used  in
experimental set ups for the value of the parameter $c=0.1$. The GP
equation we
used is strictly valid at 0 K \cite{8}. For a trap frequency $\omega \sim
2 \pi \times 100$ Hz, $\hbar \omega \sim 5$ nK. In that case from figure 1 
we find that the height of the reduced trap (\ref{st}) is around 20 nK
for a condensate at 0 K, as the scaled potential is expressed in 
units of $\hbar \omega$. However,
the condensates in the laboratory are observed often at $50-100$ nK, and
the
height of the reduced trap in the laboratory should be higher than this
value. We could not predict the minimum height for forming the BEC bound
states in different cases. However, a good conservative estimate should be 
$200-400$ nK, the correct value can only be obtained in actual
experiment. The quantitative results about critical nonlinearity for the
formation of BEC states in figure 4 and about the frequency of
oscillation of the BEC in figure 6 could be verified in the laboratory
and compared with theoretical predictions.

%\acknowledgments
\ack

The work was supported in part by the CNPq 
of Brazil.

\end{document}